\documentstyle[prl,aps,epsfig,floats]{revtex}
\title {Reply to the Comment on Low energy magnetic excitations of the Mn$_{12}$-acetate spin cluster observed by neutron scattering } 
\author{I. Mirebeau$^1$, M. Hennion$^1$, H. Casalta$^2$, H. Andres$^3$, H. U. G\"udel$^3$
 A. V. Irodova$^4$, and A. Caneschi$^5$}

\address{1 - Laboratoire L\'eon Brillouin, CEA-CNRS, CE Saclay, 91191 Gif sur Yvette, France}
\address{2 - Institut Laue Langevin BP 156 F-38042 Grenoble, France}
\address{3-Department of Chemistry, University of Bern, 3000 Bern 9, Switzerland}
\address{4 -Russian Research Center, Kurchatov Institute 123182 Moscow, Russian Federation}
\address{5 -Department of Chemistry, University of Florence, Via Maragliano 7, 50144, 
Firenze, Italy}

\date{\today, submitted to Phy. Rev. Lett.}
\begin{document}
\twocolumn[\hsize\textwidth\columnwidth\hsize\csname @twocolumnfalse\endcsname
\maketitle
\begin{abstract}
Original paper  from Mirebeau et al,  PRL 83, 628, 1999.

 Comment from W. Bao et al. (cond-mat/0008042, 2Aug 2000, submitted to PRL, updated on 1 sep 2000)

\end{abstract}
\pacs{PACS numbers: 75.50Tt, 75.10.Jm, 75.40.Gb}
]
 
 The authors of the Comment present new data and contest the value previously found for the 
coefficient of the transverse fourth order term,B$_4^4$= $\pm$ 3.0(5) 10$^{-5}$ cm$^{-1}$. They argue that the 
value B$_4^4$ = 0 could explain their data. A precise determination of this term is important to 
understand the origin of the tunneling effect in Mn$_{12}$-ac.

The new neutron data shown by Bao et al are incomplete. They do not show the raw data and 
do not specify which background was subtracted and how it was measured. A careful  
background correction is essential to determine the shapes and intensities of the low energy 
peaks accurately, and therefore to statute about the transverse term. Moreover, the authors  
show only half of the spectrum, corresponding to the negative energy transfer. A correct 
analysis should be based on the whole energy range.

The data and calculations shown in Fig 1 of the comment cannot correspond to the 
temperature of 20.7 K mentioned in the text. We have checked by two independent 
calculations that at 20.7 K the  peak 1  (at -1.24 meV) should have a lower intensity than 
the peak 2  (at -1.06 meV), due to a depopulation of the lowest energy sublevel. Putting 
the right temperature is crucial. We have reanalyzed the data of Bao et al. By inserting a 
temperature of 18 K in the calculations we correctly reproduce the first seven energy peaks, 
which is not the case in Fig 1 of the comment. Note that in Fig 1 of the comment, the calculated curve is systematically below 
the experimental points for the peaks  4, 5, and 6 and 7. This misfits hinders the effect of the 
transverse parameter B$_4^4$, namely the step-like decrease of the low energy peaks 8 and 9 with 
respect to the others. In the new Fig.1 shown below, we show calculations with several B$_4^4$ values from zero to 3 
 10$^{-5}$ cm$^{-1}$, at the temperature of 18 K. We  take the energy dependence of the peak linewidth 
into account, which was likely not done in the comment. We compare our calculations with 
the data of Bao et al. The influence of the transverse term is seen on the peaks 8 and 9, in the 
energy range -0.1, -0.3 meV. The calculation with B$_4^4$ = 0  clearly overestimates the peaks, as 
quoted in the original paper. The large statistical error and the uncertainty about the background correction 
do not allow one to choose between the values 2 10$^{-5}$ cm$^{-1}$ and 3 10$^{-5}$ cm$^{-1}$.

 Finally the new data do not contradict the previous ones. In spite of the better experimental 
resolution, they do not improve the accuracy in the determination of the transverse fourth 
order term.

\begin{figure}
\centerline{\epsfig{file=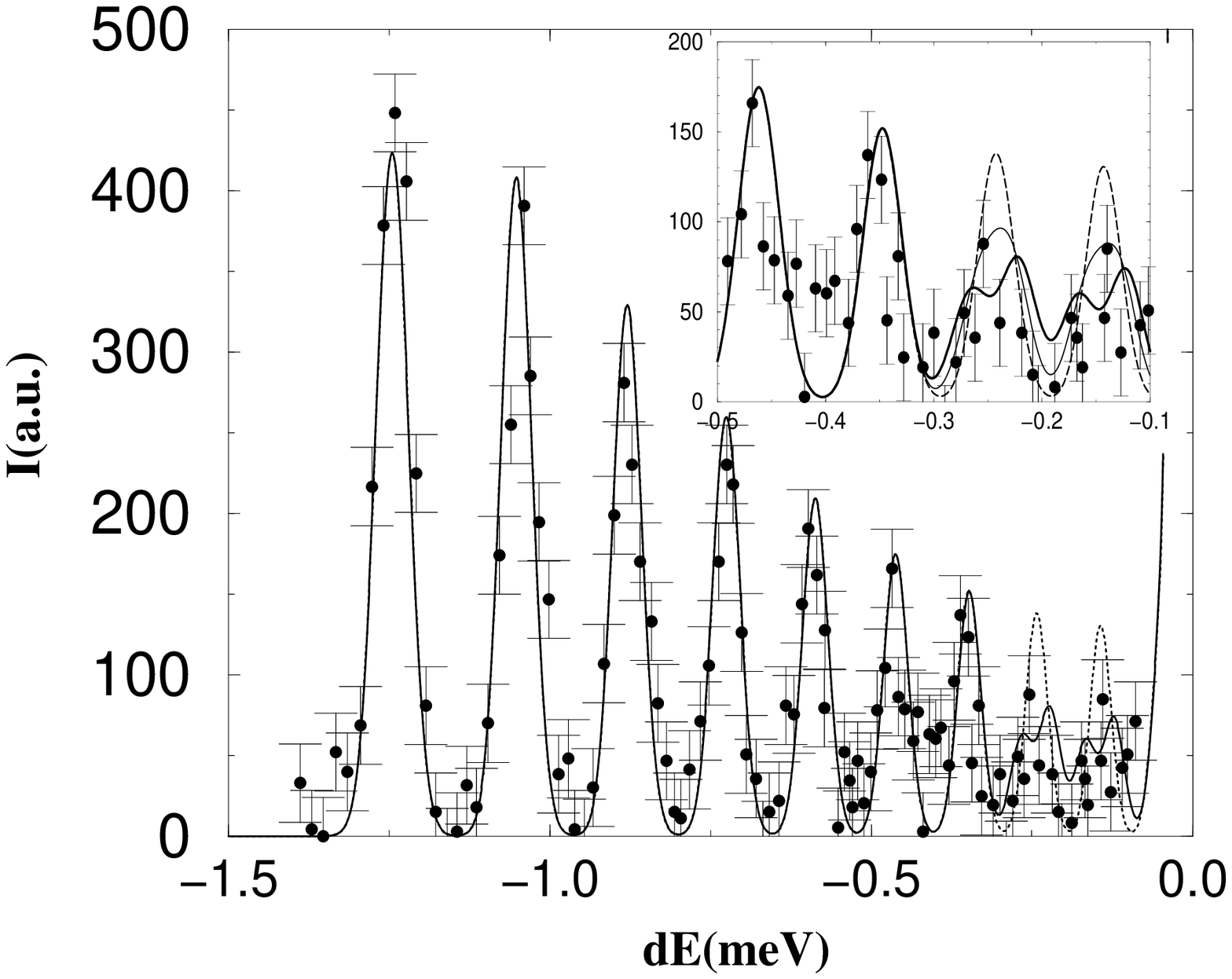,height=7cm,width=9cm}}
\caption{ Inelastic neutron scattering spectrum of Mn$_{12}$-ac. The peaks are numbered from 1 to 9 as energy decreases (peak 1 is at -1.24 meV). The data of Bao et al are compared 
with new calculations at the temperature of 18 K, for  the values B$_4^4$ = 3  10$^{-5}$ cm$^{-1}$ (solid line), 
2 10$^{-5}$ cm$^{-1}$ (thin  line) and 0 (dashed line).  In inset the low energy part of the spectrum, showing the peaks 6 to 9.}
\label{fig:38_6.8} \end{figure}

\end{document}